# Highly Tunable Band Inversion in $AB_2X_4$ ($A$=Ge, Sn, Pb; $B$=As, Sb, Bi; $X$=Se, Te) Compounds

Lin-Lin Wang

Ames Laboratory, U.S. Department of Energy, Ames, IA 50011, USA

## Abstract

Topological materials have been discovered so far largely by searching for existing compounds in crystallographic databases, but there are potentially new topological materials with desirable features that have not been synthesized. One of the desirable features is high tunability resulted from the band inversion with a very small direct band gap, which can be tuned by changes in pressure or strain to induce a topological phase transition. Using density functional theory (DFT) calculations, we have studied the septuple layered $AB_2X_4$ series compounds, where $A$=(Ge, Sn and Pb), $B$=(As, Sb and Bi) and $X$=(Se and Te). With the DFT thermodynamic stability validated by the already reported compounds in these series, we predict new stable Se compounds, which are not found in crystallographic database. Among them, we find that $GeBi_2Se_4$ and $GeSb_2Se_4$ having a small direct band gap at the Z point are very close to a strong topological insulator, which can be tuned by a moderate pressure to induce the band inversion. Importantly, the topological features with the small direct band gap are well isolated in both momentum and energy windows, which offers high tunability for studying the topological phase transition.

llw@ameslab.gov

# I. Introduction

Symmetry-protected topological states[1-4] have greatly expanded our understanding of electronic band structures in condensed matter physics and materials science. With the development of new understanding in band structure topology from symmetry-based indicators and the related approaches[5-11], symmetry-protected topological features have been classified and topological materials have been tabulated in databases. Several groups have complied catalogues[12-15] of topological materials by searching crystallographic databases among existing compounds, calculating band structures and topological index. However, there are potentially new topologically non-trivial compounds that have not been synthesized or in crystallographic databases. Even more interesting is that such new compounds may bring new tunability and push the limits in topological features. One such limit is to find a topological insulator (TI) with a band gap larger than $Bi_2Se_3$. But the other limit of a very small direct band gap is also interesting for the purpose of high tunability to induce topological phase transitions. $ZrTe_5$ is a good example sitting near the phase boundary[16, 17] between a strong and weak TI with a critical Dirac point[18] in between, which can be tuned by pressure, strain, temperature and phonons. Weyl points and many other interesting properties[19, 20] can also be induced in $ZrTe_5$. Thus, it is desirable to find more compounds sitting near the topological phase boundary to a normal insulator or another topological phase with the key features isolated in both energy and momentum windows for versatile control and manipulation.

One way to search for new topological materials is to look into the under-explored area in phase diagrams and combinations of elements. For the septuple layered $AB_2X_4$ series, where $A$=(Ge, Sn and Pb) for group IV, $B$=(As, Sb and Bi) for group V and $X$=(Se,Te) for group VI elements, the Te compounds are well explored, while Se ones are not. Eremeev et al[21] have studied several 1-2-4 compounds and found most are STIs. For example, $PbBi_2Te_4$ has been reported[22] from angular-resolved photo-electron spectroscopy (ARPES) as a STI with a band inversion at the Z point of the Brillouin zone (BZ) boundary, in contrast to that at Γ point of BZ center in the $Bi_2Te_3$ series. More recently the magnetic version in this structure-type $MnBi_2Te_4$ series[23-29] have been discovered to be an intrinsic anti-ferromagnetic TI with the few layers hosting quantum anomalous Hall and axion insulator behaviors[30, 31]. In contrast, the Se compounds have been much less studied. Only very recently, MnBi$_2$Se$_4$ thin films[32] have been grown in molecular beam epitaxy. Se-Te alloyed septuple layered compounds have also been recently synthesized[33].

In this paper, using density functional theory (DFT) calculations with different exchange-correlation (XC) functionals, we systematically calculate the thermodynamic stability, electronic band structure and topological properties of both Te and Se compounds in $AB_2X_4$ series. Out of nine different cation combinations for Te compounds in this structure, seven have been reported, except for SnAs$_2$Te$_4$ and PbAs$_2$Te$_4$, which we find are thermodynamically unstable with decomposition into two binary compounds in our DFT calculations, agreeing with experimental observations. In contrast, the only Se compounds that have been reported[34, 35] are SnBi$_2$Se$_4$ and PbBi$_2$Se$_4$, with the latter being an alloy. Using these reported compounds to establish the predictability of the DFT calculations for phase stability, we find that eight of nine Se compounds are thermodynamically stable with respect to the decomposition into the binary compounds. Although with the modified Becke-Johnson[36] (mBJ) exchange potential to correction band gap, the band inversion in Se compounds is lifted, we find that two new stable Se compounds, GeBi$_2$Se$_4$ and GeSb$_2$Se$_4$, are very close to a STI with a small direct band gap at the Z point. We predict that a moderate hydrostatic pressure, can induce the topological phase transition with the band inversion and a critical Dirac point at the Z point, such as 1.7 GPa for GeBi$_2$Se$_4$. More importantly, these topological features are well isolated in both momentum and energy windows, which offers high tunability and clean Fermi surface for studying the topological phase transition.

## II. Computational Methods

All density functional theory[37, 38] (DFT) calculations with spin-orbit coupling (SOC) have been performed with different XC functionals using a plane-wave basis set and projector augmented wave method, as implemented in the Vienna Ab-initio Simulation Package[39, 40] (VASP). Besides PBEsol[41], for van der Waals (vdW) interaction we have considered vdW density functional (vdW-DF) of optB86b[42]. We have also used the semi-empirical vdW parameter sets of D2[43], D3[44] and D3 with Becke-Johnson damping[45] (or D3BJ). We use a kinetic energy cutoff of 300 eV, $\Gamma$-centered Monkhorst-Pack[46] (12×12×4) $k$-point mesh, and a Gaussian smearing of 0.05 eV. The ionic positions and unit cell vectors are fully relaxed with the remaining absolute force on each atom is less than 0.02 eV/Å. For a few cases with available experimental lattice parameters, ionic relaxation only has also been considered for comparison to the results from full relaxation. Using maximally localized Wannier functions[47, 48], tight-binding models have been constructed to reproduce closely the band structure including SOC within 1eV of the Fermi level by using Group IV *sp*, Group V and VI *p* orbitals. The surface spectral functions have been calculated with the surface Green's function methods[49, 50] as implemented in WannierTools[51].

## III. Results and Discussion

### III-A. Structural-energy prediction

$AB_2X_4$ in the rhombohedral $GeAs_2Te_4$-type (Pearson symbol hR21) crystal structure of space group 166 ($R\bar{3}m$) and its first Brillouin zone (BZ) are shown in Fig.1(a), where *A*=(Ge, Sn and Pb), *B*=(As, Sb and Bi), and *X*=(Se and Te), respectively. It has septuple layered units in the *ab* plane coupled weakly via vdW interaction along the (001) direction. Within each septuple unit, the stacking is *X-B-X-A-X-B-X*, where A (group IV element) occupies the central, B (group V) and X (group VI) elements occupy the three outmost atomic layers. The two group VI sites are not equivalent. Group IV, V and VI elements have the nominal valence charge of +2, +3 and –2, respectively. Such structure belongs to a large group of rhombohedral and hexagonal vdW layered materials with charge balance for being narrow gap semiconductors, thermoelectrics and TIs.

The DFT-calculated formation energy ($E_f$) in eV per formula unit (eV/f.u.) are plotted for $AB_2X_4$ with respect to binary *AX* and $B_2X_3$ compounds in Fig.1(b)-(f) for different XC functionals with SOC. The ground state crystal structures of the binary *AX* and $B_2X_3$ compounds are listed in Table I. These binary compounds are also fully relaxed with the corresponding XC functionals for the $E_f$ calculation. Considering the deviation of DFT-calculated $E_f$ vs experimental values due to the limitation of XC functional, there is likely a shift of energy zero[52]. For example, in Fig.1(d) with D3+SOC, $GeBi_2Te_4$ has $E_f$=0.05 eV/f.u., but it was found stable in experiment. When the effective energy zero is shifted to that of $GeBi_2Te_4$, we find seven out of nine Te compounds are stable, except for $PbAs_2Te_4$ and $SnAs_2Te_4$, which is in a good agreement with experimental observations. The results from the five different XC functionals in Fig.1 all agree that $PbAs_2Te_4$ and SnAs$_2$Te$_4$ are the two unstable ones among the nine Te compounds. Using the criteria that the rest seven Te compounds are thermodynamically stable as found in experiment, an energy shift of about E$_f$=0.05 eV/f.u. is needed for all five XC functionals. The almost same size of the energy zero shift indicates a systematic error in PBE-based XC functionals for these compounds.

With the energy zero established, the results for Se compounds can be separated into two groups. PBEsol and D3BJ give similar results to each other that all Se compounds are stable, while D2, D3 and optB86b give all are stable except for PbAs$_2$Se$_4$. The change of E$_f$ across the series is also similar for PBEsol and D3BJ, but different from the other three XC functionals. The difference among the two groups of XC functionals also reflects in the optimized lattice constants and band gaps as we will discuss later. From Fig.1, the overall trend is that Se compounds on average are more stable than the corresponding Te ones with respect to the decomposition into binaries, except for some Se compounds with D2. Another important trend for Te compounds is that, except for Ge series, the stability increases with group V element going from As to Sb to Bi. This trend is also largely true for the Se compounds with D2, D3 and optB86b. Interestingly, we find that eight out of nine Se compounds are thermodynamically stable with respect to binary compounds, except for PbAs$_2$Se$_4$, as determined by most XC functionals used.

The good predictability of DFT can also be seen in the optimized lattice constants. The direct comparison between the DFT-optimized lattice constants $a_{DFT}$ and the nine available experimental $a_{expt}$ with the five XC functionals are plotted in Fig.2 (a)-(e) and $c_{DFT}$ vs $c_{expt}$ in Fig.2 (f)-(j), respectively. Again, PBEsol and D3BJ give very similar results of more accurate *a*, but *c* is largely underestimated. While both D3 and optB86b overestimate *a* by similar amount, the latter give better prediction in *c*. Lastly D2 slightly underestimates *a*, the *c* prediction is as good as optB86b. These differences are also directly reflected in the mean absolute error (MAE) and root mean squared error (RMSE) as listed in Table II. While PBEsol and D3BJ have the smallest MAE and RMSE for *a*, those for *c* are the largest. Even for D3+SOC, it slightly overestimates *a*, while underestimates *c*, but they are both within +/–2%, the acceptable range for DFT predictability. The experimental data are mostly for Te compounds. Notably the Se data (red dot) for SnBi$_2$Se$_4$ and PbBi$_2$Se$_4$ are in line with the Te compounds, as well as the alloyed structures (open symbols).

Together with the thermodynamic stability in Fig.1, these comparison to experimental data shows the reliable predictability of DFT for structural energy of these layered compounds. There are some differences among the five XC functionals. Noticeably in contrast to D2, D3 with geometry-dependent dispersion coefficients and $r^{-8}$ besides $r^{-6}$ terms give results very close to the vdW-DF XC functional optB86b. Then interestingly, instead of the zero-damping, D3BJ for D3 with the BJ damping gives very similar results to PEBsol for these compounds.

### III-B. Band-gap prediction

After establishing the reliable predictability of DFT structural energy for $AB_2X_4$, we next discuss the predicted band gap, band structure and topological properties. The combination of vdW-DF XC functional and mBJ exchange potential has been shown to give good results to band gaps for vdW materials[53, 54] with the latter being a meta-GGA to add orbital-dependent correction using the converged Kohn-Sham wave functions of the vdW-DF optimized crystal structures. Here we will also combine mBJ with PBEsol XC functional to calculate the band structures, as well as other semi-empirical vdW energy corrections with the correspondingly optimized crystal structures. Among the 18 compounds, four Te compounds, $GeSb_2Te_4$, $GeBi_2Te_4$, $SnSb_2Te_4$ and $PbBi_2Te_4$, have their band gap measured experimentally and were determined as STI by ARPES. For these compounds, because of the available experimental lattice constants, we will also calculate band structures with only ionic relaxations (IR), besides the full relaxation (FR) in different XC functionals without and with mBJ exchange potential. These different band gap results are listed in Table III for comparison and also plotted in Fig.3(a) and (b) for IR and FR, respectively. The open symbols are for the results with mBJ correction.

As listed in Table III, for $GeSb_2Te_4$, a *p*-doped STI with a small band gap of 0.1 eV as determined[55] by ARPES, IR structures give a trivial narrow-gap semiconductor with a band gap from 0.00 to 0.07 eV, except for optB86b giving it as a STI with a gap of 0.02 eV. With mBJ, the band gap for the trivial narrow-gap semiconductor increases by 0.07 eV. The band inversion for STI is removed in optB86b with mBJ, although the resulting band gap size is also 0.02 eV. All five XC functionals in IR with mBJ give $GeSb_2Te_4$ as trivial. In a distinct contrast, all five XC functionals with FR structures give it as a STI agreeing with experiment and the calculated band gap ranges from 0.05 to 0.11 eV. With mBJ, the band gap for FR structures all decreases, but it remains as a STI for all five XC functionals. These opposite changes for GeSb$_2$Te$_4$ in IR and FR without and with mBJ can be seen as the opposite relative positions of the open and filled symbols at $\Delta_{expt}$=0.1 eV for GeSb$_2$Te$_4$ in Fig.3(a) and (b), respectively. These differences are showcased by the bulk band structures calculated with D3+SOC in Fig.3(c)-(f) for GeSb$_2$Te$_4$, for which the BZ and high symmetry $k$-points are shown in Fig.1(a). The band inversion in these compounds happens near the $E_F$ at the Z point. As shown in Fig.3(c) for IR, the small gap of 0.02 eV is a direct one without band inversion at the Z point. With mBJ in Fig.3(d), the trivial direct gap in enlarged to 0.09 eV, as the lowest conduction band at the F point is pushed to higher energy and the highest valence band along L-Z is pushed to lower energy, which are expected with mBJ. In contrast, for FR in Fig.3(e), the band gap of 0.11 eV near the Z point is inverted, clearly shown by the green shade projected on the outmost Te $p$ orbitals. With mBJ, the lowest conduction band at the F point is still pushed higher and the highest valence band along L-Z is pushed lower in energy aiming to increase the band gap. But because of the band inversion, the tendency of mBJ to separate valence and conduction band needs to first reduce the band inversion region, but not necessarily the band gap itself. As seen in Fig.3(f) when compared to (e), the size of the band inversion region around the Z point becomes smaller with mBJ, but as the band inversion is not lifted, the band gap between the inverted bands actually decreases, changing from 0.11 to 0.07 eV. Thus, the correction with mBJ can either increase or decrease the band gap size, depending on the presence of band inversion or not, as well as if the gap is a direct or indirect one.

For the other three experimentally measured Te compounds as STI, GeBi$_2$Te$_4$ was confirmed[56] to have a surface Dirac point by ARPES with a bulk band gap of 0.2 eV. A later ARPES study[57] put the bulk band gap at 0.12 eV. SnSb$_2$Te$_4$ is a heavily $p$-doped STI, the ARPES study estimated[58] the bulk gap to be a minimum of 0.2 eV. Our calculated band gap agrees with the previous PBEsol calculation[59] of 0.12 eV. For PbBi$_2$Te$_4$, it has been measured[22] by ARPES as a STI with a band gap of 0.23 eV. Our calculated band gap also agrees with the previous calculation[22], giving an underestimated band gap around 0.12 eV. From the MAE and RMSE compiled in Table IV that are based on the limited data of the four Te compounds, the D3 emerges as the choice of XC functional for band structure calculation. D3 has the smallest MAE and its RMSE is also not far from the smallest. Thus, for topological band structures that will be discussed below, D3 will be used. The D3+SOC band structures with FR of the other three known STI Te compounds are plotted in Fig.4. It clearly shows that the band inversion region for the Bi compounds is larger than the Sb one due to the stronger SOC. The band inversion region is also reduced when using mBJ. The band gap can increase or decrease depending on if it is a direct gap near the Z point like $GeSb_2Te_4$ or indirect gap like these three compounds. But for all the four Te compounds, the FR structures with five different XC functional predict they are all STI, agreeing with experiment.

Ideally more experimental data from Se compounds are needed to more effectively validate the DFT prediction of band gap. From the MAE and RMSE in Table IV, PBEsol and D3BJ give good lattice constant of $a$, but the calculated band gap for FR structures are not as close to experiments as the other group of XC functionals, mostly due to the inaccurate prediction on $c$, which is also important for the band inversion at the Z point. Although IR with experimental lattice constants for band structure calculation has been regarded as reliable most of time, here it misses the band inversion in $GeSb_2Te_4$. This shows the sensitivity of band inversion and band gap to the change of lattice constants besides the ionic coordinates, which makes pressure an effective knob to tune topological phase transition in these compounds.

The calculated band gap for all the compounds with different XC functionals are plotted in Fig.5. These results are with FR structures without and with mBJ potential to correct the band gap. The Te compounds mostly have an indirect gap, while the Se ones mostly have a direct band gap at the Z point. As shown in Fig.5, across the different compounds, the results from D3BJ are quite similar to those from PBEsol, following the same kind of similarity in the predicted lattice constants and formation energy. For Te compounds, except for D2, the mBJ band gaps are very close to the one without such correction. For Se compounds, the ones with As have the largest band gap predicted with mBJ, because they are normal semiconductors without band inversion. When there is a band inversion, the mBJ needs to first overcome the band inversion before enlarge the band gap. So for compounds with band inversion, mBJ does not necessarily increase much the band gap. In contrast for trivial systems without band inversion, mBJ can open up a significantly larger band gap, which are mostly As compounds.

## III-C. Topological tunability

For topological classification of the band structures of $AB_2X_4$, due to the inversion symmetry, the Fu-Kane[60] type $Z_2$ topological index based on parity eigenvalues can be calculated. The septuple structure is known for the band inversion at the Z point with the Fu-Kane topological index of (1;111), in contrast to the band inversion at the Γ point in the quintuple-layered $Bi_2Te_3$-type. The band inversion is between group VI $p_z$ and group V $p_x$ and $p_y$ orbitals. From our calculations, most Te compounds with Sb and Bi are STI, while most As ones are not, except for $SnAs_2Te_4$, but it is not thermodynamically stable. These results show the importance of SOC for band inversion, because As is the lightest in group V. Without mBJ correction, most Se compounds except for the ones with As are calculated as STI and some have sizable band gap and some have a very small gap. With mBJ correction, all Se compounds calculated in D2, D3 and optB86b become topologically trivial semiconductors, but importantly these band gaps are quite small, which can be tuned by pressure.

For example, we predict two thermodynamically stable Se compounds, $GeBi_2Se_4$ and $GeSb_2Se_4$, not found in crystallographic database, are narrow-gap semiconductors based on mBJ, but close to the topological phase transition to STI. Without mBJ, $GeBi_2Se_4$ and $GeSb_2Se_4$ are STI having a very small direct band gaps of 0.01 and 0.04 eV, and the bulk band structures are shown in Fig.6(b) and (c), respectively. With mBJ, the band inversion is lifted and the gap is increased to 0.22 and 0.26 eV (Fig.6(e) and (f)). Unlike most Te compounds having an indirect band gap, such as $SnBi_2Te_4$ in Fig.6(a) and (d), the band gap for these Se compounds is direct at the Z point. Away from the Z point, the top valence and bottom conduction bands are well separated (Fig.6(e) and (f)), thus these two Se compounds have their tunable band inversion region isolated in both momentum and energy windows, which is a desirable feature for experimental detection and control of the topological properties. For these Se compounds, especially $GeBi_2Se_4$, the small band gap means it is very close to the critical point of topological phase transition between a STI and normal insulator with a critical Dirac point at the Z point, which can be tuned by pressure, strain, phonon and temperature. Previously ZrTe$_5$ is a prototype of such highly tunable topological material. Here we predict that GeBi$_2$Se$_4$ and GeSb$_2$Se$_4$ are also materials with high tunability in band structure topology. The sensitivity of the band dispersion to pressure here is similar to that in $AE$Cd$_2$As$_2$ series[61] for inducing a Dirac point.

To demonstrate the high tunability of these narrow-gap semiconductors that are close to STI, the surface spectral functions of GeBi$_2$Se$_4$ (001) have been calculated with different hydrostatic pressure ($P$) up to 2.0 GPa as shown in Fig.7. First, the Wannier charge center (WCC) evolution on the $k_z$=0.5 plane in Fig.7(a) shows the nontrivial band topology for GeBi$_2$Se$_4$ under 2.0 GPa with the band inversion at the Z point. As plotted in Fig.7(b), the spectral function has topological surface states enveloping the bulk conduction bands (see Fig.1(a) for the hexagonal surface BZ) and converging toward the $\bar{\Gamma}$ point, which is well isolated in both momentum and energy windows. Figure 7(c) shows the in-plane spin texture from the upper surface Dirac cone enveloping the bulk conduction bands at energy as high as E$_F$+0.2 eV. Because the band inversion is only at the Z point, when projected on (001), the non-trivial surface Dirac point is expected to appear at $\bar{\Gamma}$ point. After zoomed in Fig.7(d), indeed there is only one surface Dirac point at $\bar{\Gamma}$ point with a small bulk band gap. With smaller pressure, the band gap is first closed in Fig.7(e) and then reopened in Fig.7(f) showing a topological phase transition. Less than $P$=1.6 GPa as in Fig.7(f), the band inversion is not present, which is confirmed by the surface states only attaching to the bulk conduction bands for a normal insulator. Importantly, at $P$=1.7 GPa in Fig.7(e), a critical bulk Dirac point emerges at the Z point with both the surface states and Dirac point projection converging at the $\bar{\Gamma}$ point. Thus, we propose that the high tunability of the topological phase transition in GeBi$_2$Se$_4$ and GeSb$_2$Se$_4$ can potentially be explored with pressure, strain and coherent phonons as found in pump-probe experiments for ZrTe$_5$. We also emphasize the need to have experimentally measured band gaps for Se compounds to validate the different DFT-calculated results.

## IV. Conclusion

In conclusion, using density functional theory (DFT) calculations, we have studied the septuple layered $AB_2X_4$ series compounds with $A$=(Ge, Sn and Pb), $B$=(As, Sb and Bi) and $X$=(Se and Te). Unlike the seven Te compounds, there are only two Se compounds been reported in these series. Using these reported compounds to validate the thermodynamic stability from DFT calculations with different exchange-correlation functionals, we predict new stable Se compounds that are not found in crystallographic database. Interestingly, we find $GeBi_2Se_4$ and $GeSb_2Se_4$ with a small direct band gap are close to a topological phase transition to strong topological insulators with a band inversion at the Z point, which can be induced by a moderate hydrostatic pressure. Additionally, the topological features with the small direct band gap in these compounds are well isolated in both momentum and energy windows, which offers high tunability for studying the topological phase transition. Using $GeBi_2Se_4$ as the example, we show that under a small hydrostatic pressure about 1.7 GPa, the band gap can be closed and then inverted with a critical Dirac point appearing at the Z point.

## Acknowledgements

This work at Ames Laboratory was supported by the U.S. Department of Energy, Office of Science, Basic Energy Sciences, Materials Sciences and Engineering Division. The Ames Laboratory is operated for the U.S. Department of Energy by Iowa State University under Contract No. DE-AC02-07CH11358.

| *AX* | Ge | Sn | Pb |
|---|---|---|---|
| Se | oP8, 62, GeS | oP8, 62, GeS | cF8, 225, NaCl |
| Te | hR6, 160, GeTe | cF8, 225, NaCl | cF8, 225, NaCl |

| $B_2X_3$ | As | Sb | Bi |
|---|---|---|---|
| Se | mP20, 14, $As_2S_3$ | oP20, 62, $Sb_2Se_3$ | hR15, 166, $Bi_2Te_3$ |
| Te | hR15, 166, $Bi_2Te_3$ | hR15, 166, $Bi_2Te_3$ | hR15, 166, $Bi_2Te_3$ |

Table I. Binary *AX* and $B_2X_3$ compounds in their ground state crystal structure as listed in the sequence of Pearson symbol, space group and structural type.

| MAE/RMSE (Å) | *a* | *c* |
|---|---|---|
| FR-PBEsol | **0.014/0.017** | 0.826/0.923 |
| FR-D2 | 0.027/0.030 | 0.450/0.495 |
| FR-D3 | 0.055/0.059 | 0.606/0.694 |
| FR-D3BJ | 0.018/0.020 | 0.821/0.935 |
| FR-optB86b | 0.043/0.046 | **0.378/0.490** |

Table II. Mean absolute error (MAE) and root mean squared error (RMSE) for the fully-relaxed nine compounds with different XC functionals in Fig.2. Numbers in bold highlight the minimum MAE and RMSE.

| Δ (eV) | GeSb$_2$Te$_4$ | GeBi$_2$Te$_4$ | SnSb$_2$Te$_4$ | PbBi$_2$Te$_4$ |
|---|---|---|---|---|
| expt | 0.1 | 0.12, 0.2 | 0.2 | 0.23 |
| IR-PBEsol/mBJ | *0.07/0.14* | 0.08/0.06 | 0.14/0.09 | 0.12/0.15 |
| IR-D2/mBJ | *0.06/0.13* | 0.06/0.04 | 0.10/0.02 | 0.13/0.13 |
| IR-D3/mBJ | *0.02/0.09* | 0.09/0.07 | 0.16/0.12 | 0.12/0.16 |
| IR-D3BJ/mBJ | *0.00/0.07* | 0.07/0.07 | 0.16/0.12 | 0.12/0.15 |
| IR-optB86b/mBJ | 0.02/*0.02* | 0.09/0.07 | 0.18/0.13 | 0.12/0.16 |
| FR-PBEsol/mBJ | 0.08/0.03 | 0.09/0.06 | 0.09/0.16 | 0.10/0.13 |
| FR-D2/mBJ | 0.05/0.02 | 0.07/0.03 | 0.07/0.00 | 0.13/0.15 |
| FR-D3/mBJ | 0.11/0.07 | 0.10/0.07 | 0.14/0.13 | 0.12/0.12 |
| FR-D3BJ/mBJ | 0.07/0.03 | 0.07/0.06 | 0.10/0.16 | 0.12/0.15 |
| FR-optB86b/mBJ | 0.09/0.07 | 0.07/0.08 | 0.13/0.13 | 0.13/0.14 |

Table III. Band gaps of the four compounds with experimental data compared to the calculated results with ionic relaxation (IR) in experimental lattice constants and fully relaxed (FR) structures using different XC functionals without and with mBJ exchange potential. The numbers in italics indicate trivial band gap, while the others are with band inversion for STI.

| MAE/RMSE (eV) | IR | FR |
|---|---|---|
| PBEsol | 0.059/0.066 | 0.072/0.087 |
| PBEsol-mBJ | 0.074/0.078 | 0.068/0.071 |
| D2 | 0.076/0.080 | 0.082/0.089 |
| D2-mBJ | 0.098/0.113 | 0.112/0.123 |
| D3 | 0.066/0.074 | **0.052**/0.066 |
| D3-mBJ | **0.054/0.061** | 0.066/0.073 |
| D3BJ | 0.076/0.083 | 0.072/0.080 |
| D3BJ-mBJ | 0.061/0.065 | 0.062/0.064 |
| optB86b | 0.061/0.072 | 0.058/0.066 |
| optB86b-mBJ | 0.069/0.070 | 0.059/**0.063** |

Table IV. Mean absolute error (MAE) and root mean squared error (RMSE) for the four compounds with experimental data compared to the calculated results with structures from the ionic relaxation in experimental lattice constants and full relaxation using the different XC functionals and also mBJ as plotted in Fig.3. Numbers in bold highlight the minimum MAE and RMSE.

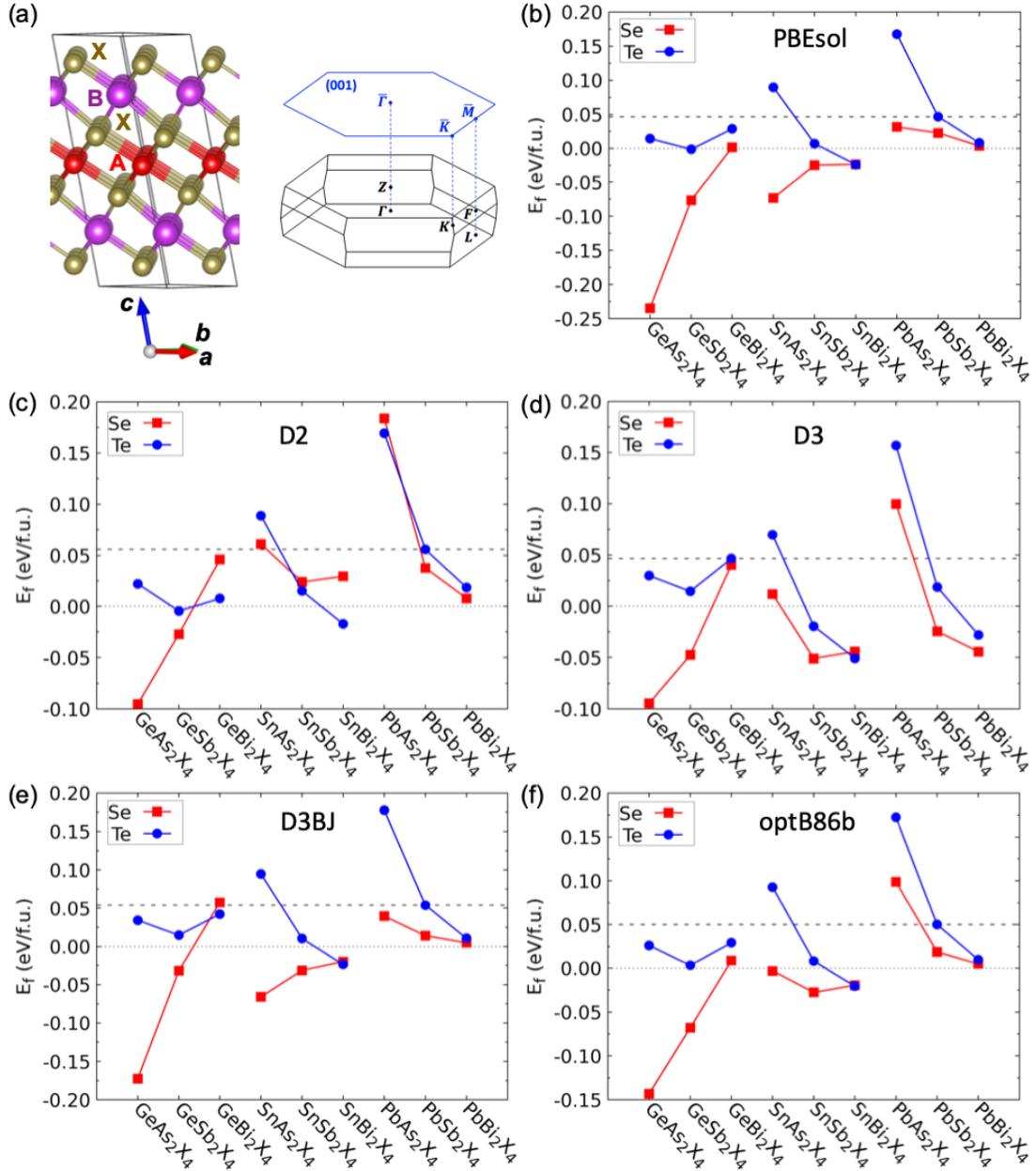

Figure 1. (a) Crystal structure and Brillouin zone (BZ) of $AB_2X_4$ in the rhombohedral GeAs$_2$Te$_4$-type (hR21) of space group 166 ($R\bar{3}m$). $A$=(Ge, Sn and Pb) are in red, $B$=(As, Sb and Bi) in purple and $X$=(Se and Te) in brown spheres. High symmetry $k$-points and those on (001) surface BZ are labeled. (b)-(f) DFT-calculated formation energy (E$_f$) with different XC functional with SOC in eV per formula unit (f.u.) of $AB_2X_4$ with respect to binary $AX$ and $B_2X_3$ compounds. The dotted and dashed lines are for energy zero and that shifted to GeBi$_2$Te$_4$ or PbSb$_2$Te$_4$ as found stable in experiment, respectively.

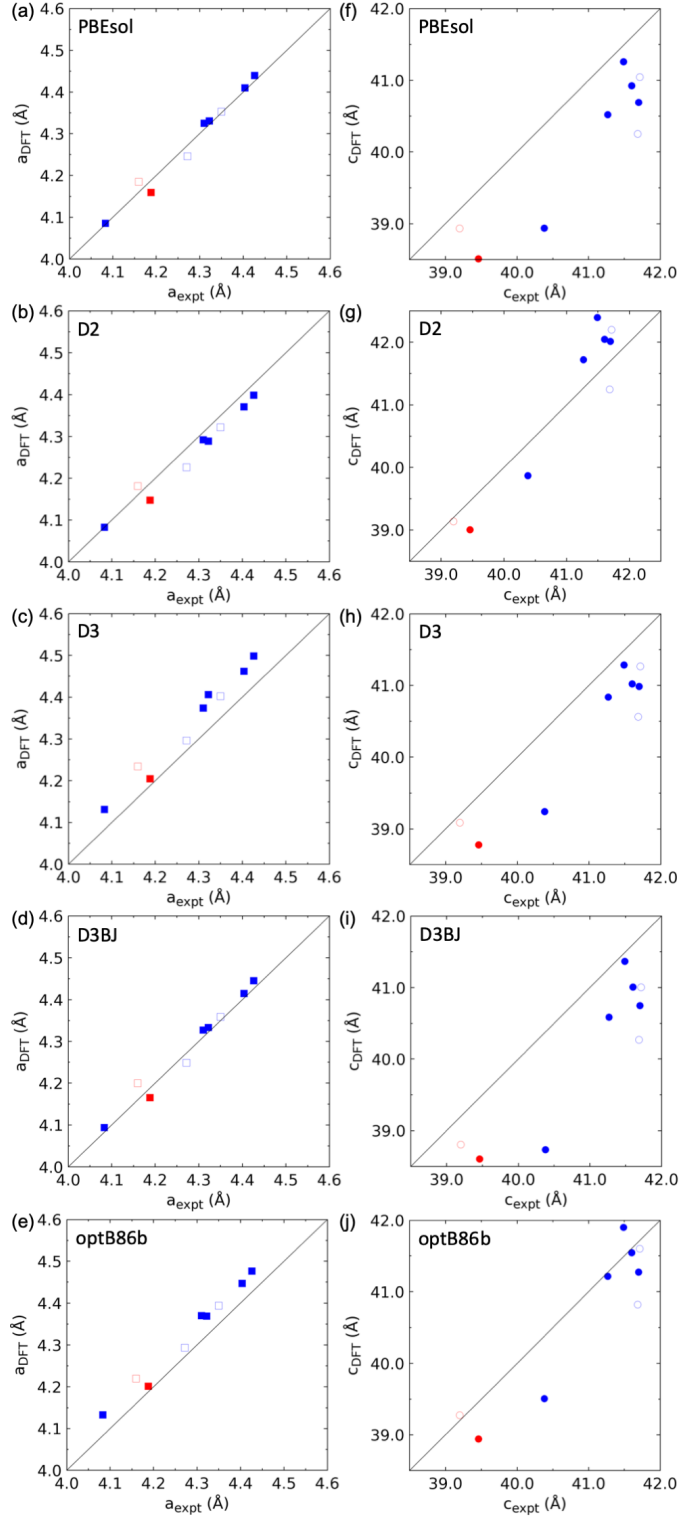

Figure 2. Comparison of the DFT-optimized lattice constants with experimental data for (a)-(e) *a* and (f)-(j) *c* using different XC functionals for nine $AB_2X_4$. The blue and red are for $AB_2Te_4$ and $AB_2Se_4$, respectively. The open symbols are reported alloys for $GeSb_2Te_4$, $PbSb_2Te_4$ and $PbBi_2Se_4$.

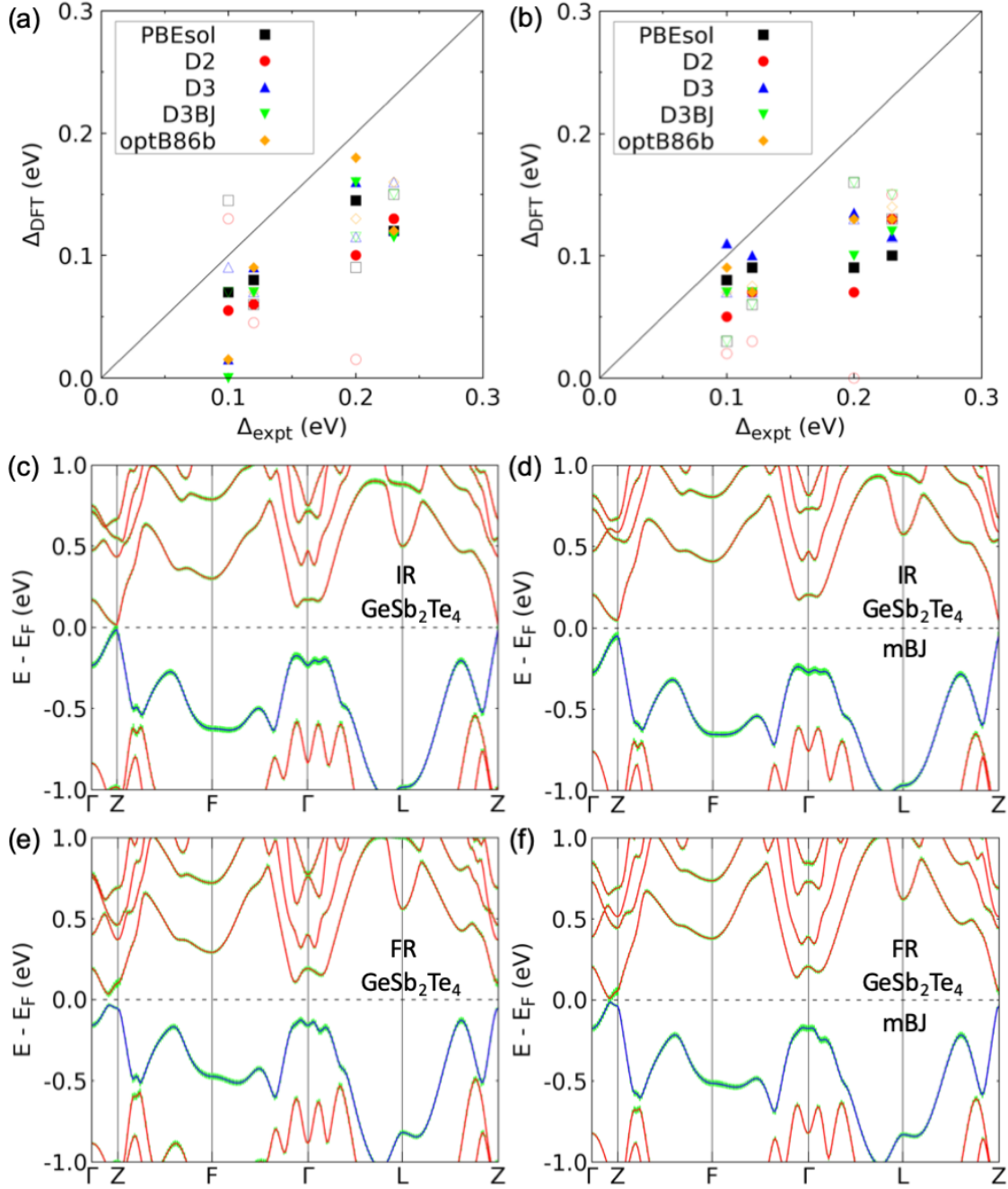

Figure 3. Comparison between the four available experimental band gap ($\Delta_{expt}$) and the calculated ones ($\Delta_{DFT}$) using (a) ionically relaxed (IR) structures with experimental lattice constants and (b) fully relaxed (FR) with DFT-optimized lattice constants. The open symbols are mBJ results for the corresponding XC functionals. Band structure of GeSb$_2$Te$_4$ calculated using D3+SOC with (c) IR structure and (d) also mBJ. Those with (e) FR and (f) also mBJ. The top valence band is in blue and the green shadow stands for the projection on Te $p$ orbitals of the outmost layer. The high symmetry $k$-points are in Fig.1(a).

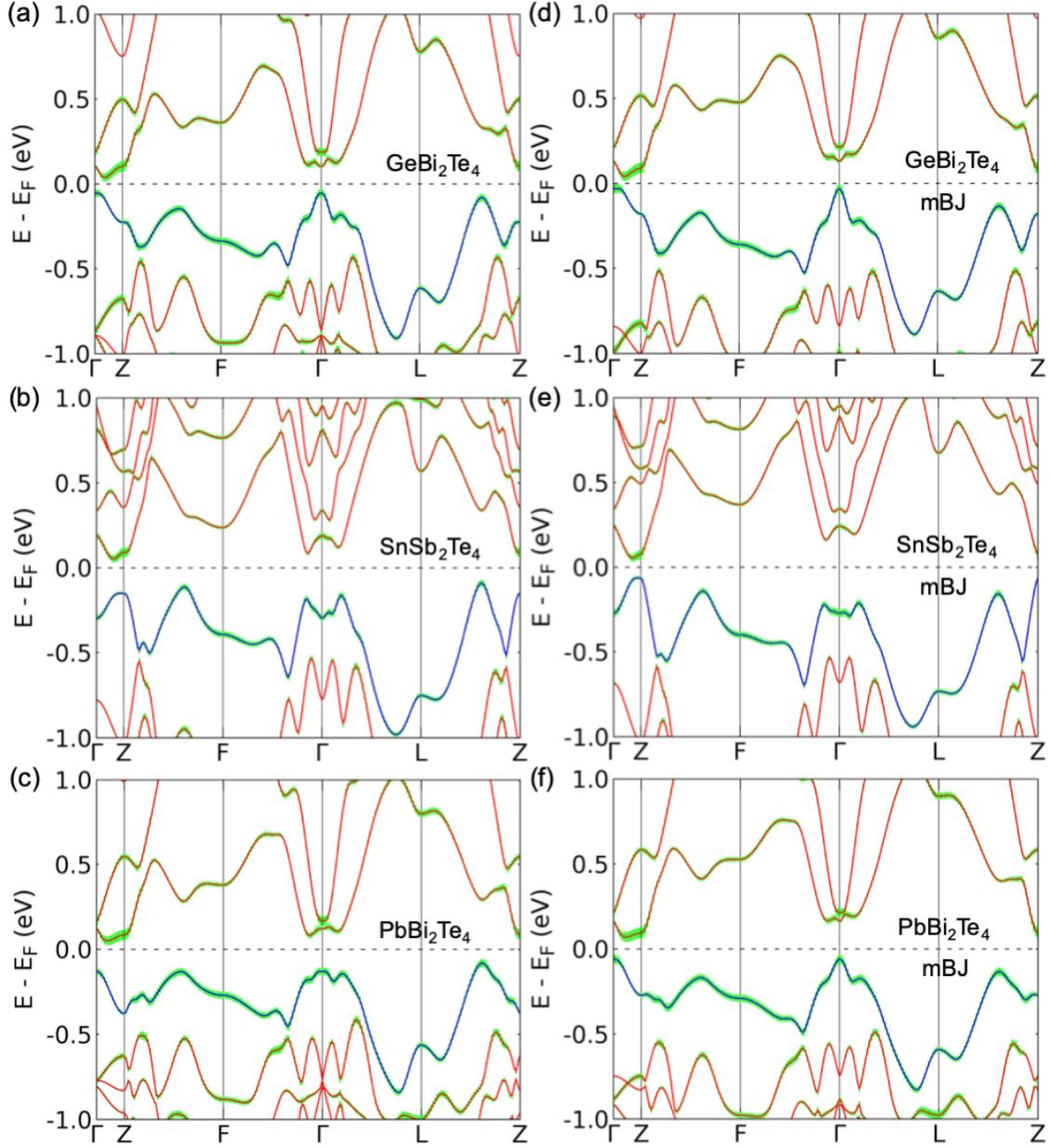

Figure 4. (a)-(c) Band structure calculated in D3+SOC with the fully relaxed crystal structures for GeBi$_2$Te$_4$, SnSb$_2$Te$_4$ and PbBi$_2$Te$_4$, respectively. (d)-(f) The corresponding band structures with mBJ. The top valence band is in blue and the green shadow stands for the projection on Te *p* orbitals of the outmost layer.

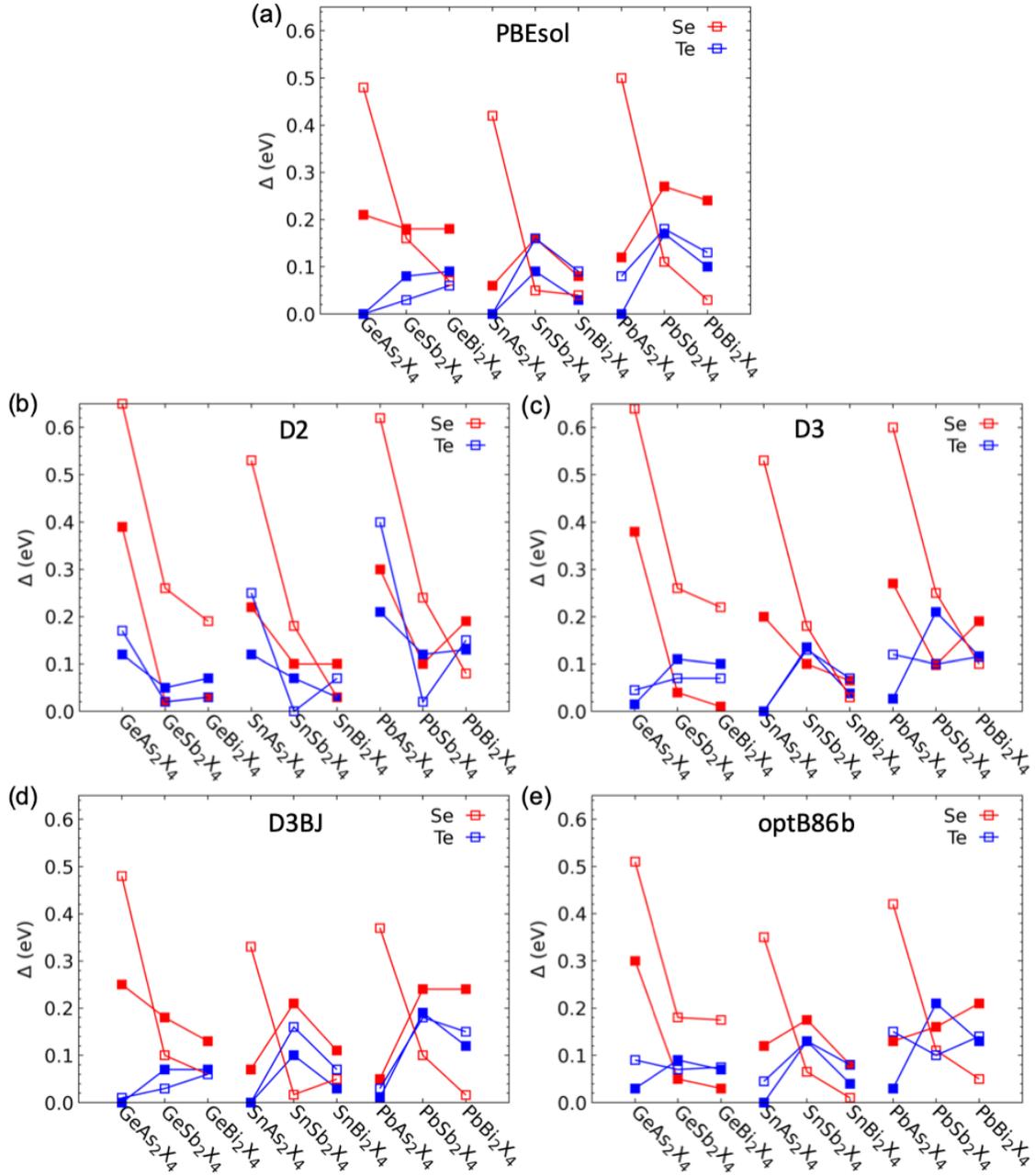

Figure 5. DFT-calculated band gap (Δ) for $AB_2X_4$. The blue and red are for $AB_2Te_4$ and $AB_2Se_4$, respectively. The solid and open symbols are for the different XC functionals and the corresponding mBJ results, respectively.

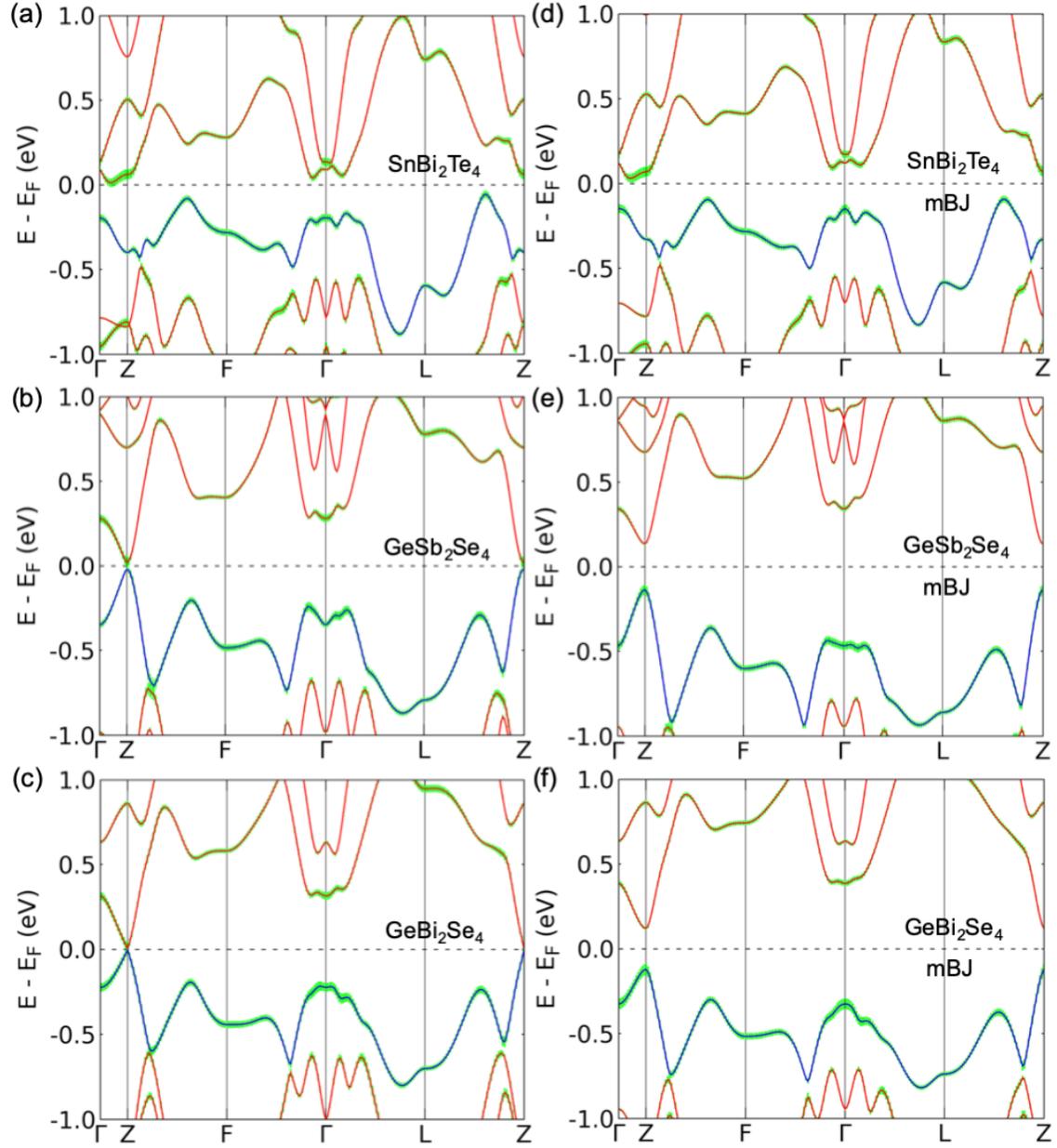

Figure 6. (a)-(c) Band structure calculated in D3+SOC with the fully relaxed crystal structures for SnBi$_2$Te$_4$, GeSb$_2$Se$_4$ and GeBi$_2$Se$_4$, respectively. (d)-(f) The corresponding band structures with mBJ. The top valence band is in blue and the green shadow stands for the projection on $X$ $p$ orbitals of the outmost layer.

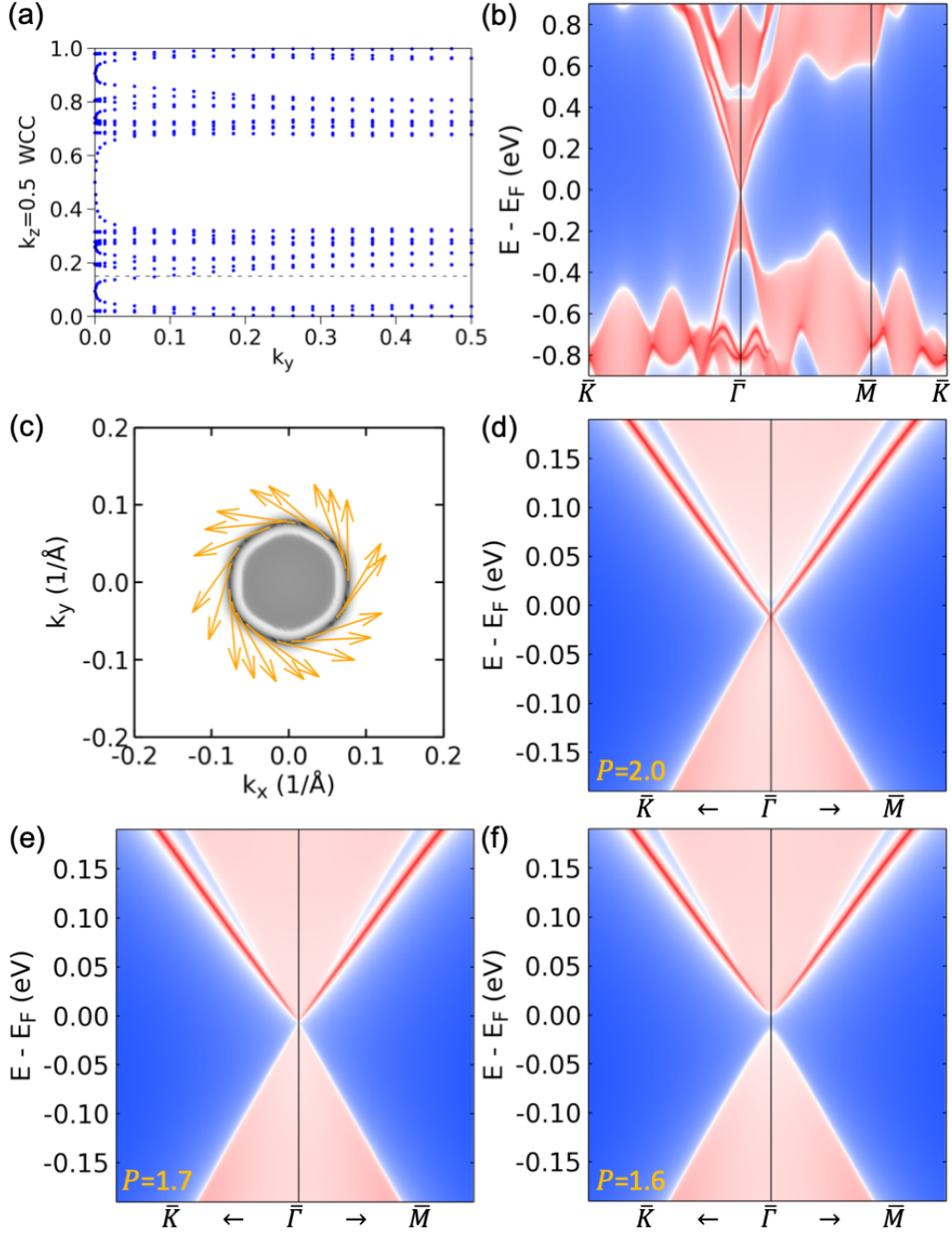

Figure 7. (a) Wannier charge center (WCC) evolution on $k_z$=0.5 plane, (b) (001) surface spectral function and (c) spin texture on 2D Fermi surface at $E_F$+0.2 eV for GeBi$_2$Se$_4$ with the hydrostatic pressure ($P$) of 2.0 GPa, showing the non-trivial topology of a STI with $Z_2$=(1;111). The high symmetry $k$-points are in Fig.1(a). The surface spectral function zoomed around $\bar{\Gamma}$ point with $P$ of (d) 2.0, (e) 1.7 and (f) 1.6 GPa, calculated with mBJ+SOC. The $P$=1.7 GPa gives a critical bulk Dirac point.